\documentclass[dvips,twoside, a4paper,12pt]{preprint}

\usepackage{times}
\usepackage{theorem}

\usepackage{mhequ}
\usepackage{mhfig}

\def\ij{ij}
\def\ii{ii}
\def\q{{\bf q}}
\def\x{{\bf x}}
\def\p{{\bf p}}

\def\SS{{\cal S}}

\def\LL{{\cal L}}
\def\HH{{\cal H}}
\def\HALF{{\textstyle{1\over 2}}}
\def\Ref#1{(\ref{#1})}
\def\real{{\bf R}}
\def\Dp{\nabla_{\negthinspace p}}
\def\Dq{\nabla_{\negthinspace q}}
\def\moy#1{\langle#1\rangle}
\def\vir{{\rm,}}
\def\vers{\rightarrow}
\def\ve{\varepsilon}

\def\vt{\vartheta}
\def\VT{\Theta}
\def\Id{{\bf 1}}
\def\Ze{{\bf 0}}

\theoremstyle{plain}
\theorembodyfont{\it}
\newtheorem{Lemma}{Lemma}
\theorembodyfont{\rm}
\newtheorem{Example}{Example}

\begin{document}

\title{Strange Heat Flux in (An)Harmonic Networks}
\author{J.-P. Eckmann${}^{\rm1,2}$ and E. Zabey${}^{\rm1}$}
\institute{%
${}^1$\,D\'epartement de Physique Th\'eorique,\\$\,{}^2$\,Section de Math\'ematiques,\\
\hphantom{${}^1$}\,Universit\'e de Gen\`eve, Switzerland}

\maketitle

\begin{abstract}
We study the heat transport in systems of coupled oscillators driven
out of equilibrium by Gaussian heat baths. 
We illustrate with a few examples that such systems can exhibit
``strange'' transport phenomena. In particular, {\em circulation} of heat flux
may appear in the steady state of a system of three oscillators only. 
This indicates that the direction of the heat fluxes can
in general not be ``guessed" from the temperatures of the heat baths. 
Although we primarily consider
harmonic couplings between the oscillators, we explain why this
strange behavior persists under weak anharmonic perturbations.
\end{abstract}

\bigskip


\bigskip
\bigskip
\bigskip

%
%
\section{Networks of oscillators}

In this note, we consider steady states of (an)harmonic oscillators
driven by heat reservoirs at different temperatures. We show, by
simple examples, that
``anything is possible'' for such physical systems: in particular,
it is basically impossible to guess in which direction energy flows. 
We will first describe the harmonic case and then argue why
the results extend to mildly anharmonic problems. \\

The setup is that of $n$ masses, all equal to 1, connected by a set of
harmonic ``springs,''  at most $n(n-1)/2$ of them. For the sake of 
simplicity, the position and velocity of each mass are chosen
to be one-dimensional.
The potential is
a function $V(q_1,q_2,\dots,q_n)$, which is given as a
positive definite quadratic form $\HALF (\q,V\q)$. 
The (Gaussian) heat baths interact
with some (at least 2) of the $n$ masses. Each mass 
is either attached to its own heat bath at temperature $T_i>0$, with friction
$\Gamma_i$, or is attached to no heat bath. 
In this case we will
say that $\Gamma_i=0$ and leave $T_i$ undefined. 
The stochastic differential
equations describing such a system are for $i=1,\ldots,n$:
\begin{equa}
d p_i  \,&=\, -(V\q)_i\,dt - \Gamma_i p_i\,dt+ 
\sqrt{2\Gamma_i T_i}\,d\omega_i(t)~\vir\\
d q_i  \,&=\, p_i\,dt~\vir
\end{equa}
where the $\omega_i(t)$ are independent Wiener processes.
It will be convenient to write the problem in matrix form. 
Let $\x=(\p,\q)$ denote the state of the $2n$ masses. The 
invariant measure of the problem (if it exists), is (up to
normalization) of the form $\exp\bigl(-\HALF (\x,Q^{-1}\x)\bigl)$ with the
$2n\times 2n$ matrix $Q$ being the solution to the {\em Lyapunov equation}
\begin{equ}[e:QA]
Q A^* + AQ \,=\,-B~\vir
\end{equ}
where
\begin{equ}
A\,=\, \left ( \matrix{ -\Gamma &-V\cr \Id& \Ze\cr}\right )~,\qquad
B\,=\, \left ( \matrix{ 2\Gamma T &\Ze\cr \Ze& \Ze\cr}\right )~.
\end{equ}
Here, $\Gamma$ and $T$ are the diagonal matrices whose elements 
are $\Gamma_i$ and $T_i$. We denote by 
$\HH=\{i:\Gamma_i\neq 0\}$ the indices of the masses in direct contact 
with a heat bath.
The following condition 
assures uniqueness of the invariant measure,
and can be easily derived from \cite{ZSb}~:
%
%
\begin{Lemma}\label{thm:span}
Consider the space $\SS$ spanned by the
vectors $\{V^ke_i,i\in \HH, k=0,\dots,n\}$, where $e_i$ denotes the 
$i^{\rm th}$ unit vector of $\real^n$. If
$\SS=\real^n$, then \Ref{e:QA} has a unique solution. Moreover, this
solution is positive definite.
\end{Lemma} 

\begin{figure}[ht]
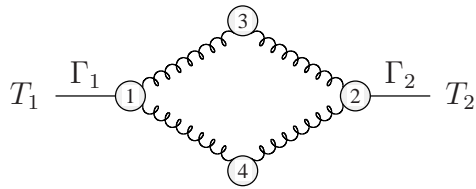

\begin{center}
\mhpastefig{los1}
\caption{Non-unique steady state}
\label{fig:losange}
\end{center}
\end{figure}

When the condition of Lemma \ref{thm:span} is not satisfied, a change
of coordinates shows that at least one mode is neither coupled to a
heat bath nor to the rest of the system. 
The simplest example  where this happens
is shown in Fig.~\ref{fig:losange} (see \cite{Diplome, Maes}). 
The masses 1 and 2 are coupled to
heat baths, while the masses 3 and 4 
are only coupled to the masses 1 and 2. All the springs have the same
coupling constant.
Writing the equations of motion, one easily checks that the variables
$q=q_3-q_4$ and $p=p_3-p_4$ evolve as an isolated harmonic 
oscillator.\\

%
%

We henceforth assume that the assumptions of 
Lemma \ref{thm:span} hold (this can be easily
verified for the examples given in the sequel).
Therefore, the steady state exists, is unique, and we
denote by $\moy{f}$ the average value of an observable $f(x)$ in this state.
For convenience, we shall write the matrix $Q$ as four $n\times n$
blocks
\begin{equ}[e:blocks]
Q\,=\,\left ( \matrix{ X &R\cr R^*& Y}\right )~,
\end{equ}
where $X$ and $Y$ are positive symmetric matrices. As a consequence of
\Ref{e:QA}, $R^*=-R$. Averages
of quadratic observables are given by the elements of the matrix $Q$, namely
$\moy{p_i\,p_j}=X_{\ij}$, $\moy{q_iq_j}=Y_{\ij}$ and $\moy{p_iq_j}=R_{\ij}$.

%
%
\section{Heat fluxes}

We briefly recall a common definition of a heat flux
between two points of the system \cite{EPRb}.
In general, the evolution of an observable 
$f$ is given by the equation $\dot f=\LL f$, where
$\LL$ is the Fokker-Planck operator, in our case 
\begin{equ}
\LL \, = \, p\cdot\Dq - q\cdot V\Dp - p\cdot\Gamma\Dp 
+ \Dp\cdot\Gamma T\Dp~.
\end{equ}
By definition we have $\moy{\LL f}=0$. 
The energy in the spring connecting points $i$ and $j$ is
$U_{\ij}=-\HALF V_{\ij}(q_i-q_j)^2$, where $V_{\ij}<0$ when the
coupling is
attractive.
In order to obtain the heat flux between these two points,
we interpret the equation $\moy{\LL U_{\ij}}=0$ as a conservation 
equation for the energy in the spring, and identify the terms
in this equation as energy fluxes.
We denote the average value of the flux from $i$ to $j$
by $\phi_{i\vers j}$, whose expression turns out to be
\begin{equ}
\phi_{i\vers j}=V_{\ij}\,\moy{p_j(q_i-q_j)}=V_{\ij}\,\moy{p_jq_i}~\vir
\end{equ}
since $\moy{p_iq_i}=R_{\ii}=0$ by antisymmetry of the matrix $R$.
For a point $i$ connected to a bath, the heat flux 
entering the system through that point, denoted by $\phi_i$,  is obtained 
similarly, leading to
\begin{equ}[e:phii]
\phi_i=\Gamma_i\,(T_i-\moy{p_i^2})~.
\end{equ}
Because of energy conservation,
the total heat flux at every point has average zero
in the steady state. In the sequel, we only consider 
average quantities and by {\em flux} we always mean {\em
average flux in the steady state}.\\

Very few results are available concerning the {\em direction}
of the heat fluxes in the system. The main one is the positivity
of the global entropy production, namely 
\begin{equ}
-\sum_{i\in\HH}\,\frac{\phi_i}{T_i} > 0~.
\end{equ}
This (strict) inequality has been proved in \cite{EPRb} 
for an anharmonic chain
between two baths at different temperatures.
Under the conditions of Lemma \ref{thm:span}, one can
easily show that it remains valid for harmonic
networks \cite{Diplome}. Because the matrix $Q$ is positive
definite, we can also conclude:
%
%
\begin{Lemma}\label{thm:Tmax}
The point(s) attached to the hottest bath(s) cannot absorb
heat from the other baths. 
The point(s) attached to the coldest bath(s) cannot
inject heat in the system.
\end{Lemma}
\noindent
{\em Proof.~}If all the temperatures are the same, say $\vt$, one easily
checks that the steady state is Gibbsian, that is 
\begin{equ}[e:Gibbs]
Q=\vt\,\pmatrix{\Id & \Ze\cr \Ze&V^{-1}}~.
\end{equ}
In this equilibrium state all the fluxes vanish and the lemma 
is trivially verified.
Consider next a system $\SS$ with at least two different temperatures,
and denote by $Q$ the solution to the corresponding Eq.\Ref{e:QA}. 
Let $T_{\rm max}$ be the temperature of the hottest heat bath(s) of $\SS$ and
$\VT>T_{\rm max}$ be an arbitrary higher temperature.
We define a system $\SS'$  as a copy of $\SS$ but whose
temperature matrix $T'$ is given by
\begin{equa}
T'_{i}=\left\{
\begin{array}{ll}
\VT-T_i>0&{\rm for\ }i\in\HH~\vir\cr
0&{\rm otherwise~.}
\end{array}\right.
\end{equa}
Let $Q'$ be the solution to Eq.\Ref{e:QA} for $\SS'$. We note that
when all the parameters but $T$ are fixed in Eq.\Ref{e:QA},
the solution $Q(T)$ is linear in $T$. Therefore $Q+Q'$ is a
Gibbsian matrix \Ref{e:Gibbs} with $\vt=\VT$, in particular
\begin{equ}
X_{\ii }+X'_{\ii }=\VT~\vir
\end{equ}
where we have used
the block notation \Ref{e:blocks} for $Q$ and $Q'$. Since both
matrices are positive-definite, we have $X_{\ii }$ and $X'_{\ii }>0$, therefore
$X_{\ii }<\VT$ for any $\VT>T_{\rm max}$, and finally $X_{\ii }\leq T_{\rm max}$.
We consider next the flux $\phi_i$ entering the system through a ``hot'' point~
$i$ for which $T_i=T_{\rm max}$. Because of Eq.\Ref{e:phii} we
have
\begin{equ}
\phi_i=\Gamma_i\,(T_{\rm max}-\moy{p_i^2})=
\Gamma_i\,(T_{\rm max}-X_{\ii })\geq 0~.
\end{equ}
The corresponding inequality for the cold point(s) is obtained by
an equivalent construction. This concludes the proof of the 
Lemma~\ref{thm:Tmax}.\\

The two results we have mentioned give some information 
on how the system of oscillators exchanges heat with the baths.
We are now interested in knowing how the flux propagates 
{\em within} the system of oscillators.
{\em The main observation of this note is that ``everything'' is
possible}, basically through superposition of elementary solutions.
Indeed, by the linearity of $Q(T)$, each heat bath can be
considered as an independent flux source, and the total flux
at any point is simply obtained by adding the contribution 
of all the baths. This is how the four examples below can be found.

%
%
\begin{Example}\label{ex:1}
A linear chain 
\end{Example}%
\begin{figure}[ht]
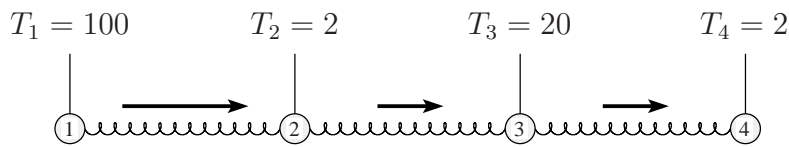

\begin{center}
\mhpastefig{example1}
\caption{The flux between 2 and 3 goes against the temperature gradient}
\label{fig:ex1}
{\scriptsize ($\Gamma_i=1, V_{12}=V_{23}=V_{34}=-1,
\phi_{1\vers2}=13.5,\phi_{2\vers3}=1.1,\phi_{3\vers4}=2.8$)}
\end{center}
\end{figure}
Consider a linear chain composed of four equal masses, each of which
is coupled to a heat bath.
In the setup of Fig.~\ref{fig:ex1}, the heat flux
is going against the (local) temperature 
gradient between the masses 2 and 3. 
Instead of defining the local temperature with the
heat bath, we can also use the local kinetic 
energy $\moy{p_i^2}$. However, we find $\moy{p_2^2}=14<18=\moy{p_3^2}$,
thus the ``backward flux'' persists.
This first example can be easily 
understood: as noted in the proof of Lemma \ref{thm:Tmax},
the matrix $Q$ solving \Ref{e:QA} is a linear function of $T$
and so are the fluxes. Thus,
we can decompose our system as the sum of two similar chains,
one with temperatures $(100,0,0,2)$ and the other with $(0,2,20,0)$.
The total flux in the middle spring still goes to the right, since
temperature $T_1$ pushes much more energy into the chain than $T_3$ does. 

%
%
\begin{Example}\label{ex:tri}
Circulation of heat
\end{Example}
\begin{figure}[ht]
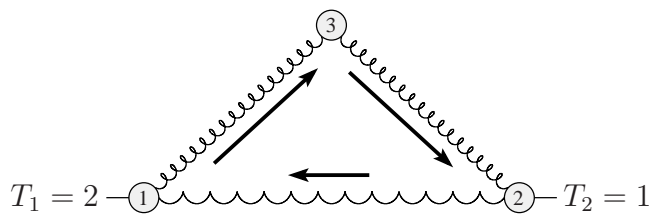

\begin{center}
\mhpastefig{example2}
\caption{A circulation of energy remains in the steady state}
\label{fig:ex2}
{\scriptsize ($\Gamma_1=\Gamma_2=1,V_{12}=-10,V_{13}=V_{23}=-20,
\phi_{1\vers3}=\phi_{2\vers3}=0.290,\phi_{1\vers2}=0.008$) }
\end{center}
\end{figure}
In this second example, the heat injected in the system
by the hot bath has two possible ``channels'' to reach the cold bath. 
What is surprising is the appearance of 
a ``backward flux'' in one of them which is not due
to excess temperature as in Example \ref{ex:1}.
As a result of this, a {\em circulation of heat} 
remains in the steady state, as shown in Fig.~\ref{fig:ex2}.
This example shows that energy fluxes between 
heat baths, as we understand them in this note,
are {\em not} similar to 
electrical currents between potentials.
Indeed, should  the arrows of Fig.~\ref{fig:ex2}
represent electrical currents, the potentials $U_i$ at points
$i=1$, $2$ and $3$ should satisfy $U_1>U_2>U_3>U_1$.
In other words, Fig.~\ref{fig:ex2} contradicts a
``Kirchoff's Law'' on current loops.
Such an example can also be constructed when the ``triangle'' is
in the center of a chain connecting two heat baths.

%
%
\begin{Example} 
Three connected heat baths with different coupling constants
\end{Example}
\begin{figure}[ht]
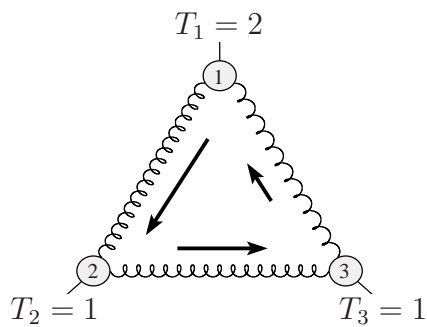

\begin{center}
\mhpastefig{example3}
\caption{Circulation in a ``fully thermalized'' triangle}
\label{fig:ex3}
{\scriptsize ($\Gamma_i=1,V_{12}=V_{23}=-45,V_{13}=-30,
\phi_{1\vers2}=0.57,\phi_{2\vers3}=0.35,\phi_{3\vers1}=0.03$) }
\end{center}
\end{figure}
The example in Fig.~\ref{fig:ex3} 
shows that the circulation of Example \ref{ex:tri}
can also be produced when all three masses are in contact with
a bath, even if two baths have the same temperature. 

%
%
\begin{Example}
A ``heat pump''
\end{Example}
\begin{figure}[ht]
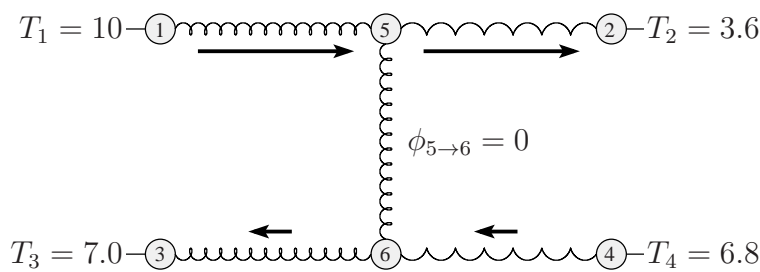

\begin{center}
\mhpastefig{example4}
\caption{The lower part of the system pumps heat from cold to hot}
\label{fig:ex4}
{\scriptsize 
($\Gamma_i=1, V_{15}=V_{56}=V_{36}=40, V_{25}=V_{46}=20,\phi_{1\vers5}=2.4,
\phi_{6\vers3}=0.2,\phi_{5\vers6}=0$)}
\end{center}
\end{figure}
In this last example, 
we construct a system that mimics a thermodynamic heat
pump. Figure \ref{fig:ex4} shows two chains of three oscillators
coupled through their middle point. 
The ends of the upper chain are
connected to the hottest $(T_1$) and the coldest ($T_2$) bath, while
the ends of the lower one are connected to intermediate temperatures ($T_3$
and $T_4$). Here again, the heat in the lower chain flows against the
temperature gradient. 
The interesting point in this example is that no energy is flowing (in average)
between the two chains ($\phi_{5\vers6}=0$). It is as if the upper chain was
acting on the lower one through {\em fluctuations} only. By slightly varying the 
temperatures, one can even obtain $\phi_{5\vers6}<0$. This last
variant is
quite different from the thermodynamics of Carnot cycles, since the subsystem in which 
heat is pumped {\em releases} energy into the pump.
We remark that Lemma \ref{thm:Tmax} prevents us from building a pump between
two baths if one of them is an extremal temperature of the system.\\

The situations described in these four examples are not a
``far from equilibrium'' behavior, because there is no such 
thing in harmonic systems. Indeed, because of the linearity of $Q(T)$,
the fluxes in the system keep the same sign when (all) temperatures are
rescaled. Moreover, fluxes are unchanged when all the temperatures
are shifted by a constant. Therefore, everything we have shown
can happen at arbitrary temperatures, and with arbitrarily small
temperature differences. We also remark that because of the 
loops of heat flux, 
there is no possible definition of local temperature that would prevent 
the flux from going against the temperature gradient.

\section{Weakly anharmonic systems}

It is well-known that the heat transport in harmonic systems does not
reproduce the usual macroscopic laws, in particular Fourier's law does
not hold \cite{RLL}.
Justification is commonly seen in the fact that
the modes of a harmonic system
are extended, causing the heat to be transported ballistically 
rather than diffusively.
Note however that phenomena described in this note
continue to hold in a weak anharmonic limit. Indeed, consider a slight
perturbation of the coupling, for instance
\begin{equ}
V_\ve(q)=\frac12\,(q,Vq)+\frac\ve4\,\sum_{i<j}\,c_{\ij}\,(q_i-q_j)^4~.
\end{equ}
The existence of a unique steady state for certain systems
with such a potential is
proved in \cite{EH}; since every point in our examples is reached in a 
simple way from a heat bath, the results of \cite{EH} generalize
to this case.
The corresponding (smooth) invariant measure 
$\rho_\ve$ is not Gaussian but still decays rapidly
at large energies.
If $\rho_\ve$ as a function of $\ve$ is sufficiently regular
around $\ve=0$, fluxes are continuous functions of this
parameter. Then, for every example we have shown, 
one can find a small enough $\ve$
so that fluxes of the perturbed system have the same direction as
those in the unperturbed system.
Although a written proof of the regularity of $\rho_\ve$ does not seem to be
available yet, this result is believed to be true \cite{SH,LS}. 
A key point is the following:
as explained in \cite{RT}, the system in the steady state 
spends most of its time below a certain energy level, with only rare
excursions to high energies. With sufficiently small $\ve$, one can make sure
that the anharmonicity is irrelevant in arbitrary long parts
of the dynamics.\\

\noindent
{\em Acknowledgments}. 
We thank M. Hairer for useful suggestions and
comments. We are also grateful to G. van Baalen, P. Collet, 
C.-A. Pillet, D. Ruelle, A. Schenkel, D. Sergi and R. Tiedra
for their helpful remarks.
This work was partially supported by the Fonds National
Suisse.

\bibliographystyle{unsrt}
\bibliography{heat}

\begin{thebibliography}{1}

\bibitem{ZSb}
M.~Zakai and J.~Snyders.
\newblock On nonnegative solutions of the equation {$AD+DA'=-C^*$}.
\newblock {\em SIAM J. Appl. Math.}, 18:704--714, 1970.

\bibitem{Diplome}
E.~Zabey.
\newblock Etats stationnaires et production d'entropie d'un syst\`eme
  harmonique hors \'equilibre.
\newblock Diploma Thesis, Universit\'e de Gen\`eve, 2001.
\newblock Unpublished.

\bibitem{Maes}
C.~Maes, K.~Netocny, and M.~Verschuere.
\newblock Heat conduction networks.
\newblock {\em J. Stat. Phys.}, 111(5):1219--1244, 2003.

\bibitem{EPRb}
J.-P. Eckmann, C.-A. Pillet, and L.~R{ey-Bellet}.
\newblock Entropy production in non-linear thermally driven hamiltonian
  systems.
\newblock {\em J. Stat. Phys.}, 95:305--331, 1999.

\bibitem{RLL}
Z.~Rieder, J.L. Lebowitz, and E.~Lieb.
\newblock Properties of a harmonic crystal in a stationary nonequilibrium
  state.
\newblock {\em J. Math. Phys}, 8:1073--1085, 1967.

\bibitem{EH}
J.-P. Eckmann and M.~Hairer.
\newblock Non-equilibrium statistical mechanics of strongly anharmonic chains
  of oscillators.
\newblock {\em Commun. Math. Phys.}, 212:105--164, 2000.

\bibitem{SH}
A.~Schenkel and M.~Hairer.
\newblock Private communications.

\bibitem{LS}
R.~Lefevre and A.~Schenkel.
\newblock Perturbative analysis of anharmonic chains of oscillators out of
  equilibrium.
\newblock {\em Preprint (mp-arc)}, 2003.

\bibitem{RT}
L.~R{ey-Bellet} and L.~E. Thomas.
\newblock Exponential convergence to non-equilibrium stationary states in
  classical statistical mechanics.
\newblock {\em Commun. Math. Phys.}, 225:305--329, 2002.

\end{thebibliography}

\end{document}